\documentclass[prl,aps,twocolumn,superscriptaddress,showpacs,showkeys,amsmath,amssymb,floatfix]{revtex4}
\usepackage[colorlinks=true,citecolor=blue,filecolor=blue,linkcolor=blue,urlcolor=blue,pdftex]{hyperref}

\usepackage{xspace} 
\usepackage[usenames,dvipsnames]{color}
\usepackage{booktabs,graphicx,mathrsfs,verbatim,amsmath,units,soul}
\usepackage{ulem}
\usepackage{xspace} 
\usepackage{upgreek}
\usepackage{gensymb}
\usepackage{tabularx}

\newcommand{\rntwo}{${}^{222}$Rn\xspace}
\newcommand{\rnzero}{${}^{220}$Rn\xspace}
\newcommand{\krbkg}{${}^{85}$Kr\xspace}
\newcommand{\krm}{${}^{83\mathrm{m}}$Kr\xspace}
\newcommand{\ambe}{${}^{241}$AmBe\xspace}

\newcommand{\cSone}{$\mathrm{cS1}$\xspace}
\newcommand{\cStwo}{$\mathrm{cS2}$\xspace}
\newcommand{\cStwob}{$\mathrm{cS2_b}$\xspace}

\newcommand{\dru}{events/(kg\,$\times$\,day\,$\times\,\mathrm{keV}_{\mathrm{ee}}$)}
\newcommand{\gevcsq}{GeV/c${}^2$}
\newcommand{\searchregion}{cS$1\!\in\![3, 70]\;\mathrm{PE}$, cS$2_b\!\in\![50, 8000]\;\mathrm{PE}$~search region}

\newcommand{\num}[1]{#1}

\makeatletter
\newcommand{\romenum}[1]{(\romannumeral #1)}
\makeatother

\newcommand{\bologna}{\affiliation{Department of Physics and Astrophysics, University of Bologna and INFN-Bologna, 40126 Bologna, Italy}}

\newcommand{\chicago}{\affiliation{Department of Physics \& Kavli Institute for Cosmological Physics, University of Chicago, Chicago, IL 60637, USA}}

\newcommand{\coimbra}{\affiliation{LIBPhys, Department of Physics, University of Coimbra, 3004-516 Coimbra, Portugal}}

\newcommand{\columbia}{\affiliation{Physics Department, Columbia University, New York, NY 10027, USA}}

\newcommand{\lngs}{\affiliation{INFN-Laboratori Nazionali del Gran Sasso and Gran Sasso Science Institute, 67100 L'Aquila, Italy}}

\newcommand{\mainz}{\affiliation{Institut f\"ur Physik \& Exzellenzcluster PRISMA, Johannes Gutenberg-Universit\"at Mainz, 55099 Mainz, Germany}}

\newcommand{\heidelberg}{\affiliation{Max-Planck-Institut f\"ur Kernphysik, 69117 Heidelberg, Germany}}

\newcommand{\munster}{\affiliation{Institut f\"ur Kernphysik, Westf\"alische Wilhelms-Universit\"at M\"unster, 48149 M\"unster, Germany}}

\newcommand{\nikhef}{\affiliation{Nikhef and the University of Amsterdam, Science Park, 1098XG Amsterdam, Netherlands}}

\newcommand{\nyuad}{\affiliation{New York University Abu Dhabi, Abu Dhabi, United Arab Emirates}}

\newcommand{\purdue}{\affiliation{Department of Physics and Astronomy, Purdue University, West Lafayette, IN 47907, USA}}

\newcommand{\rpi}{\affiliation{Department of Physics, Applied Physics and Astronomy, Rensselaer Polytechnic Institute, Troy, NY 12180, USA}}

\newcommand{\rice}{\affiliation{Department of Physics and Astronomy, Rice University, Houston, TX 77005, USA}}

\newcommand{\stockholm}{\affiliation{Oskar Klein Centre, Department of Physics, Stockholm University, AlbaNova, Stockholm SE-10691, Sweden}}

\newcommand{\subatech}{\affiliation{SUBATECH, IMT Atlantique, CNRS/IN2P3, Universit\'e de Nantes, Nantes 44307, France}}

\newcommand{\torino}{\affiliation{INFN-Torino and Osservatorio Astrofisico di Torino, 10125 Torino, Italy}}

\newcommand{\ucla}{\affiliation{Physics \& Astronomy Department, University of California, Los Angeles, CA 90095, USA}}

\newcommand{\ucsd}{\affiliation{Department of Physics, University of California, San Diego, CA 92093, USA}}

\newcommand{\wis}{\affiliation{Department of Particle Physics and Astrophysics, Weizmann Institute of Science, Rehovot 7610001, Israel}}

\newcommand{\zurich}{\affiliation{Physik-Institut, University of Zurich, 8057  Zurich, Switzerland}}

\newcommand{\paris}{\affiliation{LPNHE, Universit\'{e} Pierre et Marie Curie, Universit\'{e} Paris Diderot, CNRS/IN2P3, Paris 75252, France}}

\newcommand{\freiburg}{\affiliation{Physikalisches Institut, Universit\"at Freiburg, 79104 Freiburg, Germany}}

\begin{document}

\title{First Dark Matter Search Results from the XENON1T Experiment}

\author{E.~Aprile}\columbia
\author{J.~Aalbers}\email[]{jaalbers@nikhef.nl}\nikhef
\author{F.~Agostini}\lngs\bologna
\author{M.~Alfonsi}\mainz
\author{F.~D.~Amaro}\coimbra
\author{M.~Anthony}\columbia
\author{F.~Arneodo}\nyuad
\author{P.~Barrow}\zurich
\author{L.~Baudis}\zurich
\author{B.~Bauermeister}\stockholm
% ...
\collaboration{XENON Collaboration}\email[]{xenon@lngs.infn.it}\noaffiliation
\date{\today} 

\begin{abstract}
We report the first dark matter search results from XENON1T, a $\sim$2000-kg-target-mass dual-phase (liquid-gas) xenon time projection chamber in operation at 
the Laboratori Nazionali del Gran Sasso 
in Italy and the first ton-scale detector of this kind.  
The blinded search used \num{34.2} live days of data acquired between November 2016 and January 2017. 
Inside the \num{(1042$\pm$12)}~kg fiducial mass and in the [5, 40] $\mathrm{keV}_{\mathrm{nr}}$ energy range of interest for WIMP dark matter searches, the electronic recoil background was $(1.93 \pm 0.25) \times 10^{-4}$ \dru, the lowest ever achieved in such a dark matter detector. 
A profile likelihood analysis shows that the data is consistent with the background-only hypothesis.
We derive the most stringent exclusion limits on the spin-independent WIMP-nucleon interaction cross section for WIMP masses above 10~\gevcsq, with a minimum of 7.7~\num{$\times 10^{-47}$}~cm${}^2$ for 35-\gevcsq~WIMPs at 90\% confidence level. 

\end{abstract}



\maketitle

{\color{white}{$\;\;\;\;\;$ $\;\;\;\;\;$ $\;\;\;\;\;$ $\;\;\;\;\;$ $\;\;\;\;\;$ $\;\;\;\;\;$ $\;\;\;\;\;$ $\;\;\;\;\;$ $\;\;\;\;\;$ $\;\;\;\;\;$ $\;\;\;\;\;$ $\;\;\;\;\;$ $\;\;\;\;\;$ $\;\;\;\;\;$ $\;\;\;\;\;$ $\;\;\;\;\;$ $\;\;\;\;\;$ $\;\;\;\;\;$ $\;\;\;\;\;$ $\;\;\;\;\;$ $\;\;\;\;\;$ $\;\;\;\;\;$ $\;\;\;\;\;$ $\;\;\;\;\;$ $\;\;\;\;\;$ $\;\;\;\;\;$ $\;\;\;\;\;$ $\;\;\;\;\;$ $\;\;\;\;\;$ $\;\;\;\;\;$ $\;\;\;\;\;$ $\;\;\;\;\;$ $\;\;\;\;\;$ $\;\;\;\;\;$ $\;\;\;\;\;$ $\;\;\;\;\;$ $\;\;\;\;\;$ $\;\;\;\;\;$ }}

Modern cosmology precisely describes observational data from the galactic to cosmological scale with the $\Lambda$ cold dark matter model  \cite{Dodelson:2003ft,Kolb:1990vq}. 
This model requires a nonrelativistic nonbaryonic component of the Universe called \textit{dark matter}, with an energy density of $\Omega_c h^2 = $ \num{$0.1197 \pm 0.0022$} 
as measured by Planck 
\cite{planck_cosmo_params}. Theories beyond the Standard Model of particle physics (\textit{e.g.}, supersymmetry \cite{100gevwimp}) often attribute this energy density to weakly interacting massive particles (WIMPs) that may be detectable by underground detectors \cite{wimp,Undagoitia:2015gya}. 

The XENON1T experiment is designed primarily for detecting nuclear recoils (NRs) from WIMP-nucleus scattering, continuing the XENON program \cite{xenon10instrumentation,xenon100instrumentation} that employs dual-phase (liquid-gas) xenon time projection chambers (TPCs) \cite{xenon100instrumentation,xenon1t_mc_paper}. With a total mass of $\sim$3200~kg of ultra-pure liquid xenon --  more than two orders of magnitude larger than the initial detector of the XENON project \cite{xenon10instrumentation} -- XENON1T is the first detector  of such scale realized to date.
It is located at the Laboratori Nazionali del Gran Sasso (LNGS) in Italy, at an average depth of \num{3600} m water equivalent. The approximately 97-cm long by 96-cm wide cylindrical TPC encloses \num{$(2004\pm5)$} kg of liquid xenon (LXe), while another \num{$\sim$1200} kg provides additional shielding. 
The TPC is mounted at the center of a 9.6-m diameter, 10-m tall water tank to shield it from ambient radioactivity. An adjacent service building houses the xenon storage, cryogenics plant, data acquisition, and slow control system. The water tank is mounted with 84 photomultiplier tubes (PMTs) as part of a Cherenkov muon veto \cite{mvpaper}.
The TPC is instrumented with 248 3"~Hamamatsu R11410-21 PMTs arranged in two arrays above and below the LXe target \cite{pmt,pmt2}.
Interactions in the target produce scintillation photons (S1) and ionization electrons.
The electrons drift in a $(116.7\pm7.5)$~V/cm electric field towards the liquid-gas interface at the top of the TPC. They are extracted into the gas by an electric field \num{$E_\text{gas}>10$ kV/cm} where, via electroluminescence, they produce a proportional scintillation signal (S2). This charge-to-light amplification allows for the detection of single electrons~\cite{zeplinse, xenon100se}. 
The ratio of the S2 to S1 signals is determined by both the ratio of ionization to excitation in the initial interaction and subsequent partial recombination of the ionization, with lower S2/S1 for NR signals than electronic recoils (ERs) from $\gamma$ and $\beta$ radiation.

Here we report on \num{34.2~live~days} of blinded dark matter search data from the first science run of the experiment. The run started on November 22, 2016, and ended on January 18, 2017, when an earthquake temporarily interrupted detector operations.
The detector's temperature, pressure, and liquid level remained stable at \num{$(177.08 \pm 0.04)$ K}, \num{$(1.934 \pm 0.001)$\,bar}, and \num{(2.5 $\pm$ 0.2)\,mm} respectively, where the liquid level was measured above the grounded electrode separating the drift and extraction field regions.
While the PMT high voltage remained stable during the run, \num{27 PMTs} were turned off for the dark matter search and \num{8} were masked in the analysis due to low single-photoelectron (PE) detection efficiency. The PMT response was calibrated periodically using pulsed LED data \cite{richardledpaper}.
The xenon was continuously purified in the gas phase through hot metal getters, leading to an increase in the electron lifetime from \num{350} to \num{500}~$\mu$s, with an average of \num{452~$\mu$s}; \num{673~$\mu$s} is the drift time over the length of the TPC. 
Using cryogenic distillation \cite{column}, the ${}^{\mathrm{nat}}\mathrm{Kr}$ concentration in the LXe was reduced while the TPC was in operation, from \num{(2.60 $\pm$ 0.05)~ppt} [mol/mol] at the beginning of the science run to \num{(0.36 $\pm$ 0.06)~ppt} one month after the end of the science run, as measured by rare-gas mass spectrometry \cite{rgms} on samples extracted from the detector.
The ${}^{214}$Pb event rate was 
$(0.8-1.9)\times 10^{-4}$
 \dru~in the low-energy range of interest for WIMP searches, where the bounds are set using \textit{in-situ} $\alpha$-spectroscopy on ${}^{218}$Po and ${}^{214}$Po. 
The \rntwo concentration was reduced by $\sim$20\% relative to the equilibrium value using the krypton distillation column in inverse mode \cite{xenon100rnremoval}.

The data acquisition (DAQ) system continuously recorded individual PMT signals. The  efficiency for recording single-PE pulses was $\num{92}\%$ on average during the science run, and stable to within \num{2\%}. 
A software trigger analyzed the PMT pulses in real-time, allowing for continuous monitoring of the PMTs. The trigger detected S2s larger than 200~PE with $\num{\>99\%}$ efficiency, and saved 1~ms before and after these to ensure small S1s were captured. 
An analog-sum waveform was separately digitized together with a signal recording when any of the digitizers were inhibited. The average DAQ live time was 92\% during the science run.

\begin{figure}
\includegraphics[width=\columnwidth]{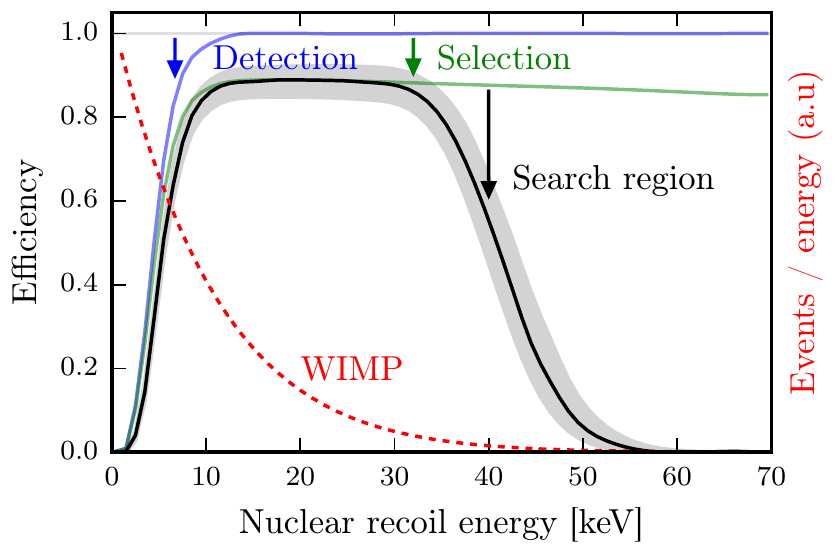}
\centering
\caption{\label{fig:efficiencies}NR detection efficiency in the fiducial mass at successive analysis stages as a function of recoil energy. At low energy, the detection efficiency (blue) dominates. 
At 20~keV, the efficiency is \num{89\%} primarily due to event selection losses (green). At high energies, the effect of restricting our data to the search region described in the text (black) is dominant. The black line is our final NR efficiency, with uncertainties shown in gray. 
The NR energy spectrum shape of a 50-\gevcsq~WIMP (in a.u.) is shown in red for reference.}
\end{figure} 

Physical signals are reconstructed from raw data by finding photon hits in each PMT channel, then clustering and classifying groups of hits as S1 or S2  using the {\sc Pax} software. For S1s, we require that hits from \num{three or more} PMTs occur within \num{50~ns}.
To tune the signal reconstruction algorithms and compute their efficiency for detecting NRs -- shown in blue in Fig.~\ref{fig:efficiencies} -- we used a Monte Carlo code that reproduces the shapes of S1s and S2s as determined by the interaction physics, light propagation, and detector-electronics chain. This was validated against \krm and \rnzero calibration data. 

The interaction position is reconstructed from the top-array PMT hit pattern of the S2 (for the transverse position) and the time difference between S1 and S2 (for depth).  The S2 transverse position is given by maximizing a likelihood based on an optical simulation of the photons produced in the S2 amplification region. The simulation-derived transverse resolution is $\sim$2~cm at our S2 analysis threshold of 200~PE (uncorrected). The interaction position is corrected for drift field nonuniformities derived from a finite element simulation, which is validated using \krm calibration data. 
We correct S2s for electron losses during drift, and both S1s and S2s for spatial variations of 
up to 30\% and 15\%, respectively, inferred from \krm calibration data. These spatial variations are mostly due to geometric light collection effects.
The resulting corrected quantities are called \cSone and \cStwo. As the bottom PMT array has a more homogeneous response to S2 light than the top, this analysis uses \cStwob, a quantity similar to \cStwo based on the S2 signal seen only by the bottom PMTs.

\begin{figure}[t!]
\centering
\includegraphics[width=\columnwidth]{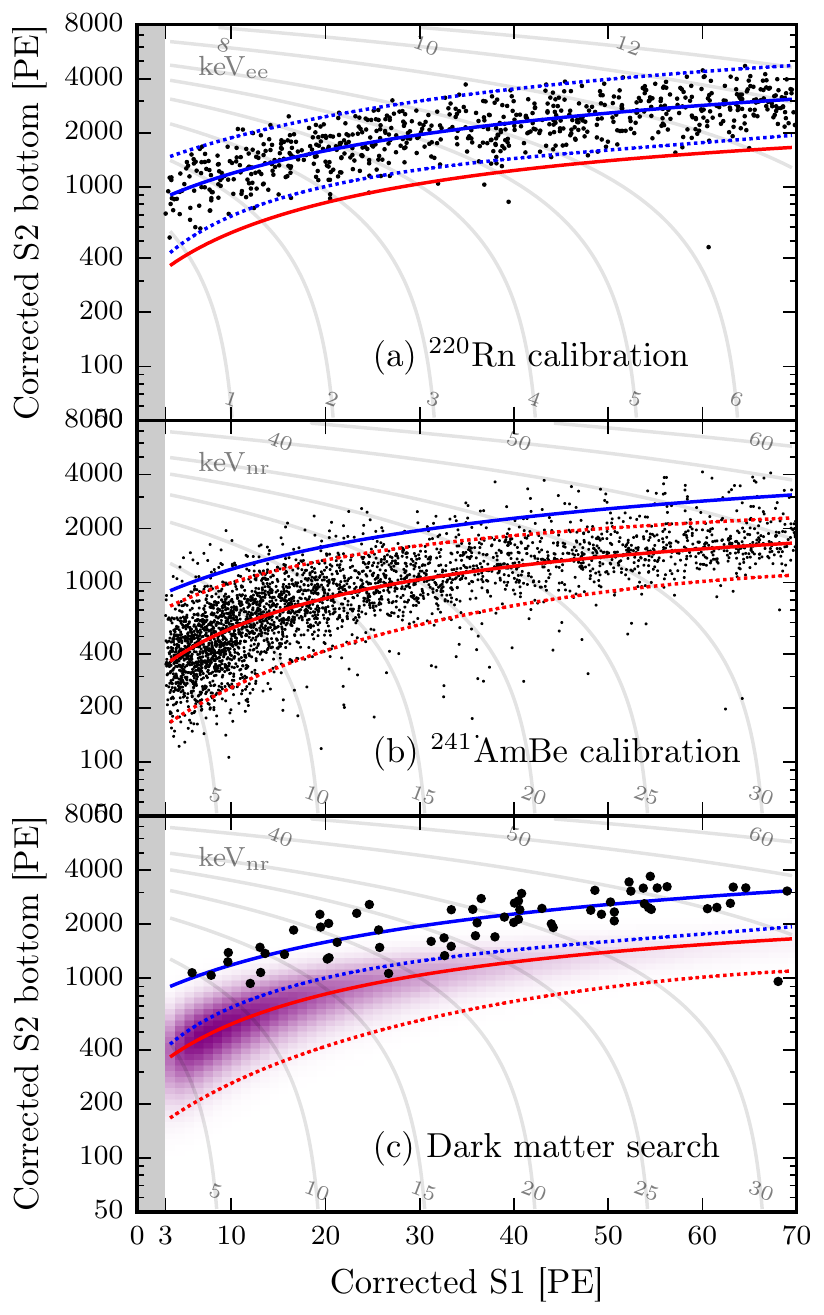}
\caption{\label{fig:data_vs_models} Observed data in \cStwob vs.~\cSone for (a) \rnzero ER calibration, (b) \ambe NR calibration, and (c) the 34.2-day dark matter search. Lines indicate the median (solid) and $\pm 2 \sigma$ (dotted) quantiles of simulated event distributions (with the simulation fitted to calibration data). Red lines show NR (fitted to \ambe) and blue ER (fitted to \rnzero). In (c), the purple distribution indicates the signal model of a 50\,\gevcsq~WIMP. Thin gray lines and labels indicate contours of constant combined energy scale in $\mathrm{keV}$ for ER (a) and NR (b, c). Data below \cSone = 3 PE (gray region) is not in our analysis region of interest and shown only for completeness.}
\end{figure}

To calibrate XENON1T, we acquired \num{3.0}~days of data with \rnzero injected into the LXe (for low-energy ERs), \num{3.3}~days with \krm injected into the LXe (for the spatial response) and \num{16.3}~days with an external \ambe source (for low-energy NRs).
The data from the \rnzero~\cite{rn220} and \ambe calibrations is shown in Fig.~\ref{fig:data_vs_models} (a) and (b), respectively.
Following the method described in~\cite{Dahl:thesis} with a $W$-value of 13.7~eV, we extracted the photon gain $g_1 = $~\num{(0.144 $\pm$ 0.007)}~PE per photon and electron gain $g_2 = $~\num{(11.5 $\pm$ 0.8)}~PE (in the bottom array, 2.86 times lower than if both arrays are used) per electron in the fiducial mass by fitting the anti-correlation of \cStwob and \cSone for signals with known energy from \krm (41.5~keV), ${}^{60} \mathrm{Co}$ from detector materials (1.173 and 1.332~MeV), and from decays of metastable ${}^{131\mathrm{m}} \mathrm{Xe}$ (164 keV) and ${}^{129\mathrm{m}} \mathrm{Xe}$ (236~keV) produced during the \ambe calibration. 
The \cSone and \cStwob yields are stable in time within \num{0.77\%} and \num{$1.2\%$} respectively, as determined by \krm calibrations. 

WIMPs are expected to induce low-energy single-scatter NRs. Events that are not single scatters in the LXe are removed by several event-selection cuts: 
\romenum{1} a single S2 above 200 PE must be present and any other S2s must be compatible with single electrons from photoionization of impurities or delayed extraction;
\romenum{2} an event must not closely follow a high-energy event (\textit{e.g.}, within 8~ms after a $3\times10^5$~PE  S2), which can cause long tails of single electrons;
\romenum{3} the S2 signal's duration must be consistent with the depth of the interaction as inferred from the drift time;
\romenum{4} the S1 and S2 hit patterns must be consistent with the reconstructed position at which these signals were produced;
\romenum{5} no more than 300~PE of uncorrelated single electrons and PMT dark counts must appear in the region before the S2.
Single scatter NR events within the [5, 40] $\mathrm{keV}_{\mathrm{nr}}$ energy range pass these selections with \num{\textgreater $82\%$} probability, as determined using simulated events or control samples derived from calibration, and shown in green in Fig.~\ref{fig:efficiencies}.
 
The dark matter search uses a cylindrical \num{(1042$\pm$12)~kg} fiducial mass, which was defined before unblinding using the reconstructed spatial distribution of ERs in the dark matter search data and the energy distribution of ERs from \rnzero. We restrict the search to cS$1\!\in\![3, 70]\;\mathrm{PE}$ and cS$2_b\!\in\![50, 8000]\;\mathrm{PE}$, which causes little additional loss of WIMP signals, as shown in black in Fig.~\ref{fig:efficiencies}.

\begin{figure}[t]
\centering
\includegraphics[width=\columnwidth]{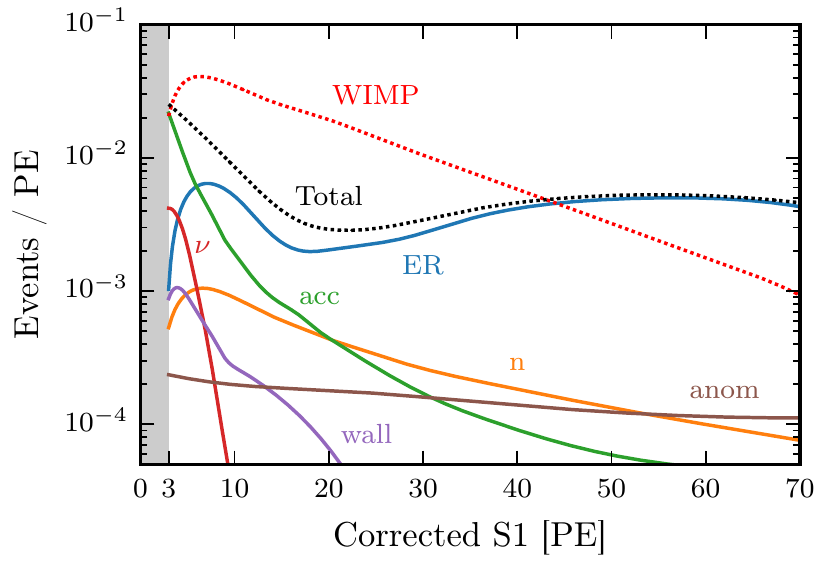}
\caption{\label{fig:bgmodel} Background model in the fiducial mass in a reference region between the NR median and $-2 \sigma$ quantile in \cStwob, projected onto \cSone. Solid lines show that the expected number of events from individual components listed in Table \ref{table:bgmodel}; the labels match the abbreviations shown in the table. 
The dotted black line \textit{Total} shows the total background model, the dotted red line \textit{WIMP} shows an $m = 50$~\gevcsq, $\sigma = 10^{-46} \mathrm{cm}^2$ WIMP signal for comparison.}
\end{figure}

Table \ref{table:bgmodel} lists the six sources of background we consider inside the fiducial mass and inside the search region.
For illustration, we also list the expected rate in a reference region between the NR median and $-2 \sigma$ quantile in \cStwob (\textit{i.e.},~between the red lines in Fig.~\ref{fig:data_vs_models}c), for which Fig.~\ref{fig:bgmodel} shows the background model projected onto \cSone.
This reference region would contain about half of the WIMP candidate events, while excluding 99.6\% of the ER background. The WIMP search likelihood analysis uses the full search region. Below we describe each background component in more detail: all event rates are understood to be inside the fiducial mass and the full search region.

\begin{table}[t]
\caption{Expected number of events for each background component in the fiducial mass; in the full  \searchregion~and in a reference region between the NR median and $-2 \sigma$ quantile in \cStwob. Uncertainties $<$0.005 events are omitted. The ER rate is unconstrained in the likelihood; for illustration, we list the best-fit values to the data in parentheses.} 
\centering
\newcolumntype{R}{>{\raggedleft\arraybackslash}X}
\bgroup
\def\arraystretch{1.2}
\begin{tabularx}{\columnwidth}{l R R}
\hline \hline
 & Full & Reference \\
\hline 
Electronic recoils (\textit{ER}) &
    (62 $\pm$ 8) &
    (0.26$\substack{+0.11 \\ -0.07}$) \\

Radiogenic neutrons ($n$) &
    0.05 $\pm$ 0.01 &
    0.02 \\

CNNS ($\nu$) &
    0.02 &
    0.01 \\

Accidental coincidences (\textit{acc}) &
    0.22 $\pm$ 0.01 &
    0.06 \\

Wall leakage (\textit{wall}) &
    0.5 $\pm$ 0.3 &
    0.01 \\

Anomalous (\textit{anom}) &
    0.10$\substack{+0.10 \\ -0.07}$ &
    0.01 $\pm$ 0.01 \\

\hline

Total background &
    63 $\pm$ 8 &
    0.36$\substack{+0.11 \\ -0.07}$ \\
    
50~\gevcsq, $10^{-46} \mathrm{cm}^{2}$ WIMP &
    1.66 $\pm$ 0.01 &
    0.82 $\pm$ 0.06 \\

\hline \hline
\vspace{0.001ex}
\end{tabularx}
\egroup
\raggedright
\label{table:bgmodel}
\end{table}

First, our background model includes ERs, primarily from $\beta$ decays of \krbkg and the intrinsic \rntwo-progeny  ${}^{214}$Pb, which cause a flat energy spectrum in the energy range of interest \cite{xenon1t_mc_paper}. 
The ER background model is based on a simulation of the detector response. We use a model similar to \cite{LUXH3paper} to convert the energy deposition from ERs into scintillation photons and ionization electrons, which we fit to \rnzero calibration data in (\cSone, \cStwob) space 
(Fig.~\ref{fig:data_vs_models}a). 

The best-fit photon yield and recombination fluctuations are comparable to those of \cite{LUXH3paper}. The model accounts for uncertainties of $g_1$, $g_2$, spatial variations of the S1 and S2 light-collection efficiencies, the electron-extraction efficiency, reconstruction and event-selection efficiency, and time dependence of the electron lifetime. The rate of ERs is not constrained in the likelihood analysis, even though we have independent concentration measurements for ${}^{214}$Pb and \krbkg, since the most stringent constraint comes from the search data itself. 

Second and third, our background model includes two sources of NRs: radiogenic neutrons contribute (0.05$\pm$0.01) events, and coherent neutrino-nucleus scattering (CNNS) $\sim$0.02 events. 
Cosmogenically produced neutrons are estimated to contribute $\mathcal{O}(10^{-3})$ events even without muon-veto tagging.
The NR background model is built from a detector response simulation that shares the same detector parameters and associated systematic uncertainties as the ER background model above. The main difference is the energy-conversion model, where we use the model and parametrization from NEST \cite{nest}. We obtain the XENON1T response to NRs by fitting the \ambe calibration data (Fig.~\ref{fig:data_vs_models}b) with the light and charge yields from \cite{nest} as priors. Our NR response model is therefore constrained by the global fit of external data.  It is also used to predict the WIMP signal models in (\cSone, \cStwob) space. The S1 detection efficiency, which is responsible for our low-energy threshold, is consistent with its prior (0.7$\sigma$).

Fourth, accidental coincidences of uncorrelated S1s and S2s are expected to contribute \num{$(0.22 \pm 0.01)$} background events. We estimated their rate and (\cSone, \cStwob) distribution using isolated S1 and S2 signals, which are observed to be at (0.78 $\pm$ 0.01) Hz and (3.23 $\pm$ 0.03) mHz, respectively, before applying S2-selections. The effect of our event selection on the accidental coincidence rate is included, similar to~\cite{xenon100combined}.
Isolated S1s may arise from interactions in regions of the detector with poor charge collection, such as below the cathode, suppressing an associated \cStwo signal. Isolated S2s might arise from photoionization at the electrodes, regions with poor light collection, or from delayed extraction \cite{extraction}. Most accidental events are expected at low \cSone and at lower \cStwob than typical NRs. 

Fifth, inward-reconstructed events from near the TPC's PTFE wall are expected to contribute \num{$(0.5 \pm 0.3)$} events, with the rate and (\cSone, \cStwob) spectrum extrapolated from events outside the fiducial mass. Most of these events would appear at unusually low \cStwob due to charge losses near the wall.
The inward reconstruction is due to limited position reconstruction resolution, especially limited for small S2s, near the 5 (out of 36) top PMTs in the outermost ring that are unavailable in this analysis.

Sixth and last, we add a small uniform background in the $(\mathrm{cS1}, \log \mathrm{cS2_b})$ space for ER events with an anomalous \cStwob. 
Such \textit{anomalous leakage} beyond accidental coincidences has been observed in XENON100~\cite{xenon100combined}, and one such event is seen in the \rnzero calibration data (Fig.~\ref{fig:data_vs_models}a).
If these were not \rnzero-induced events, their rate would scale with exposure and we would see numerous such events in the WIMP search data. We do not observe this, and therefore assume their rate is
proportional to the ER rate, at 0.10$\substack{+0.10 \\ -0.07}$ events based on the outliers observed in the \rnzero calibration data.
The physical origin of these events is under investigation.

The WIMP search data in a predefined signal box was blinded (99\% of ERs were accessible) until the event selection and the fiducial mass boundaries were finalized. We performed a staged unblinding, starting with an exposure of 4~live~days distributed evenly throughout the search period. No changes to either the event selection or background types were made at any stage.

\begin{figure}[t!]
\centering
\includegraphics[width=\columnwidth]{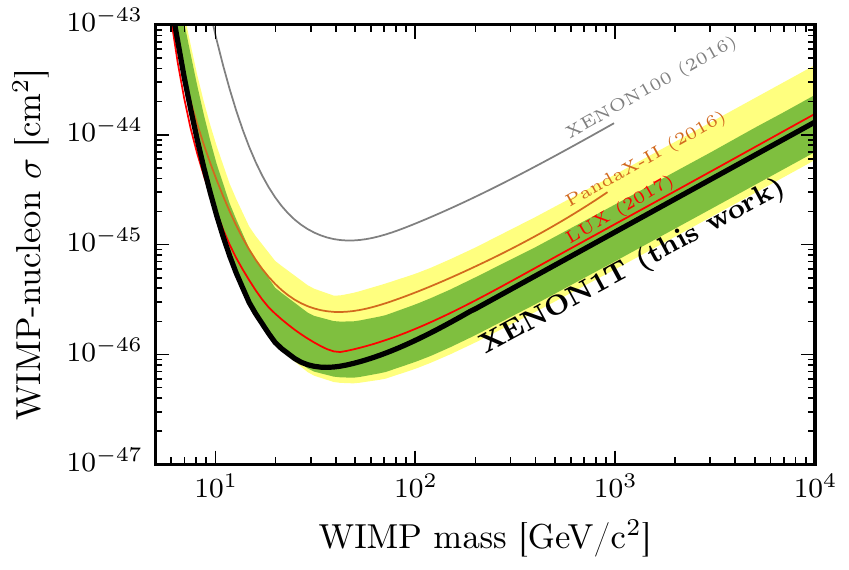}
\caption{\label{fig:blah5_limit} The spin-independent WIMP-nucleon cross section limits as a function of WIMP mass at 90\% confidence level (black) for this run of XENON1T. In green and yellow are the 1- and $2 \sigma$ sensitivity bands. Results from LUX~\cite{luxcombined} (red), PandaX-II~\cite{pandax} (brown), and XENON100~\cite{xenon100combined} (gray) are shown for reference.}
\end{figure}

A total of \num{63} events in the 34.2-day dark matter search data pass the selection criteria and are within the \searchregion~used in the likelihood analysis (Fig.~\ref{fig:data_vs_models}c). None are within \num{10 ms} of a muon veto trigger. 
The data is compatible with the ER energy spectrum in \cite{xenon1t_mc_paper} and implies an ER rate of $(1.93 \pm 0.25) \times 10^{-4}$ \dru, compatible with our prediction of $(2.3 \pm 0.2) \times 10^{-4}$ \dru~\cite{xenon1t_mc_paper} updated with the lower Kr concentration measured in the current science run. This is the lowest ER background ever achieved in such a dark matter experiment.
A single event far from the bulk distribution was observed at \num{\cSone $= 68.0$~PE} in the initial 4-day unblinding stage. This appears to be a \textit{bona fide} event, though its location in $(\mathrm{cS1}, \mathrm{cS2_b})$ (see Fig.~\ref{fig:data_vs_models}c) is extreme for all WIMP signal models and background models other than anomalous leakage and accidental coincidence. One event at $\mathrm{cS1} = 26.7~\mathrm{PE}$ is at the $-2.4 \sigma$ ER quantile.

For the statistical interpretation of the results, we
use an extended unbinned profile likelihood test statistic
in (cS1, cS2b).
We propagate the uncertainties on the most significant shape parameters (two for NR, two for ER) inferred from the posteriors of the calibration fits to the likelihood. The uncertainties on the rate of each background component mentioned above are also included. The likelihood ratio distribution is approximated by its asymptotic distribution \cite{asymform}; preliminary toy Monte Carlo checks show the effect on the exclusion significance of this conventional approximation is well within the result's statistical and systematic uncertainties.
To account for mismodeling of the ER background, we also calculated the limit using the procedure in \cite{safeguard}, which yields a similar result.

The data is consistent with the background-only hypothesis.
Fig.~\ref{fig:blah5_limit} shows the 90\% confidence level upper limit on the spin-independent WIMP-nucleon cross section, power constrained at the $-1 \sigma$ level of the sensitivity band \cite{powerconstrain}. The final limit is within 10\% of the unconstrained limit for all WIMP masses. For the WIMP energy spectrum we assume a standard isothermal WIMP halo with $v_0 = 220$~km/s, $\rho_\mathrm{DM} = 0.3$~GeV$/\text{cm}^3$, $v_\mathrm{esc} = 544$ km/s, and the Helm form factor for the nuclear cross section \cite{helm}. No light and charge emission is assumed for WIMPs below 1~keV recoil energy. 
For all WIMP masses, the background-only hypothesis provides the best fit, with none of the nuisance parameters representing the uncertainties discussed above deviating appreciably from their nominal values. Our results improve upon the previously strongest spin-independent WIMP limit for masses above 10\,\gevcsq. Our strongest exclusion limit is for 35-\gevcsq~WIMPs, at $7.7 \times10^{-47} \mathrm{cm}^2$.

These first results demonstrate that XENON1T has the lowest low-energy background level ever achieved by a dark matter experiment. The sensitivity of XENON1T is the best to date above 20 \gevcsq, up to twice the LUX sensitivity above 100 \gevcsq, and continues to improve with more data. 
The experiment resumed operation shortly after the January 18, 2017 earthquake and continues to record data.

We gratefully acknowledge support from the National Science Foundation, Swiss National Science Foundation, German Ministry for Education and Research, Max Planck Gesellschaft, Deutsche Forschungsgemeinschaft, Netherlands Organisation for Scientific Research (NWO), NLeSC, Weizmann Institute of Science, I-CORE, Pazy-Vatat, Initial Training Network Invisibles (Marie Curie Actions, PITNGA-2011-289442), Fundacao para a Ciencia e a Tecnologia, Region des Pays de la Loire, Knut and Alice Wallenberg Foundation, Kavli Foundation, and Istituto Nazionale di Fisica Nucleare. Data processing is performed using infrastructures from the Open Science Grid and European Grid Initiative. We are grateful to Laboratori Nazionali del Gran Sasso for hosting and supporting the XENON project.

\end{document}